# SUPPRESSION OF INTENSITY FLUCTUATIONS IN FREE SPACE HIGH-SPEED OPTICAL COMMUNICATION BASED ON SPECTRAL ENCODING OF A PARTIALLY COHERENT BEAM

**Gennady P. Berman**

*Theoretical Division & CNLS, MS B213, Los Alamos National Laboratory,*

*Los Alamos, New Mexico 87545*

**Alan R. Bishop**

*Theory, Simulation & Computation Directorate, MS B210, Los Alamos National Laboratory,*

*Los Alamos, New Mexico 87545*

**Boris M. Chernobrod**

*Theoretical Division & CNLS,  MS B213, Los Alamos National Laboratory,*

*Los Alamos, New Mexico 87545*

**Dinh C. Nguyen**

*ISR-6 & Accelerator Physics, Los Alamos National Laboratory MS H85,*

*Los Alamos, New Mexico 87545*

**Vyacheslav N. Gorshkov**

*Theoretical Division & CNLS, MS B213, Los Alamos National Laboratory, Los Alamos, New Mexico 87545, & the Institute of Physics, National Academy of Sciences of Ukraine, Nauki Ave. 46,  Kiev-39, 03650,  MSP-65, Ukraine*

A new concept of a free-space, high-speed (Gbps) optical communication system based on spectral encoding of radiation from a broadband pulsed laser is developed. It is shown that, in combination with the use of partially coherent laser beams and a relatively slow photosensor, scintillations can be suppressed by orders of magnitude for distances of more than 10 km. We also consider the spectral encoding of radiation from a LED as a gigabit rate solution of the "last mile" problem and rapid–deployment systems for disaster recovery. © 2006 Optical Society of America

*OCIS codes:* 030.0030, 060.4510, 350.550

# 1. Introduction

Free-space optical communication (FSOC) has data rate limitations due to atmospheric turbulence. Laser beams experience three major effects under the influence of turbulence. *First*, the beam phase front is distorted by fluctuations in the refractive index, causing intensity fluctuations or scintillations. *Second*, eddies whose size is greater than the beam diameter randomly deflect the laser beam as a whole; this phenomenon is called beam wandering. *Third*, propagation through turbulent atmosphere causes the laser beam to spread more than predicted by diffraction theory. Scintillations are the most severe problem and result in a significant increase of the bit error rate (BER) and consequent degradation of the laser communication system performance. For example, a gigabit data rate communication channel can operate with BER of $10^{-9}$ over a distance not more than 2.5 km, even for clear weather.[1,2] Several approaches have been developed to mitigate the effects of turbulence on laser communication. They include: aperture averaging, partially coherent beams, adaptive optics, and array receivers. (See the detailed reviews in[1,3].) Nevertheless, scintillations continue to limit the performance of FSOC. New approaches are needed to overcome this limitation. Recently a new technique of scintillation reduction based on the utilization of partially coherent beams (beams with multiple coherent spots in their transverse section) was demonstrated.[4-6] Combining partially coherent beams with a time-averaging photodetector leads to a significant scintillation reduction with the corresponding improvement of the BER by several orders of magnitude.[6] Unfortunately, the time-averaging method cannot be applied to gigabit rate communication. The main limitation of this method is related to the requirement that the correlation time between different spatially coherent spots be shorter than the response time of the photodetector. This means that the spatial light modulator (SLM) must have an operating frequency $\nu$ higher than the bandwidth of the photodetector, corresponding to its inverse response time $\nu \gg T^{-1}$. Since the photodetector bandwidth must be higher than the data rate of the communication channel $\nu_{COM}$, $T^{-1} \gg \nu_{COM}$, the highest data rate is limited by the highest frequency of SLM $\nu \gg \nu_{COM}$. To date, the highest frequency SLMs based on multiple quantum wells (MQW) can only operate at frequencies up to tens of MHz.[7]

In the present paper we propose to extend the technique of scintillation suppression, based on time averaging of a partially coherent beam (TAPCB), to gigabit rate FSOC. Our idea is to combine

TAPCB with a spectral encoding technique. Originally spectral encoding was applied to fiber optics communication for code-division-multiple-access.[8] In this method, information is encoded in the form of amplitude modulation of the spectral components of the laser pulse which has a broad spectrum. For long-distance communication, the broad-spectrum light source could be a Ti: sapphire laser. For short-distance communication it could be an LED as well. Each pulse or sequence of pulses (depending on the averaging response time of the photosensor) can contain kilobits of data. If the pulse repetition rate is about 1 MHz, then the transmitted data rate is gigabits per second. SLMs based on MQW technology with a frame rate of several MHz are now available.[7]

**2. Scintillations Reduction Due to Time Averaging of a Partially Coherent Beam**

It is well-established that for long distances the scintillation index of plane and spherical waves propagating through the atmospheric turbulence asymptotically tends to unity.[9] For an initially partially coherent beam, the asymptotic behavior depends on the relation between the correlation time of the source and the response time of the photodetector. If the average correlation time of two different coherent spots in the beam's cross section is shorter than the response time of the photodetector, then the scintillation index asymptotically tends to zero.[10-13] If the correlation time of the coherent spots is longer than the response time of the photodetector, then the scintillation index asymptotically tends to unity.[13,14] As was shown in [13], these properties of a partially coherent beam can be easily explained if we assume that the scintillations at the photodetector follow Gaussian statistics. Indeed, if the coherence radius, $r_c$, of the initial beam is significantly smaller than the beam radius, $r_0$, the process of propagation of the laser beam can be considered as the independent propagation of a large number of coherent beams. Consequently, the intensity fluctuations of each coherent region caused by atmospheric turbulence are statistically independent. With increasing the propagation distance, the individual coherent spots overlap due to diffraction effects. According to the Central Limit Theorem, the intensity, which is the result of the contributions of a large number of independent regions, has a normal statistical distribution. The suppression of scintillations in the signal measurements is strictly due to the unique properties of the Gaussian statistics. The fluctuations in the signal generated by a photodetector with slow response time are proportional to the following integral over light intensity absorbed during the response time:

$$\langle i(t)i(t)\rangle - \langle i^2(t)\rangle \sim \int_0^\infty dt_1 \int_0^\infty dt_2 \exp(-|t-t_1|/T - |t-t_2|/T)[\langle I(t_1)I(t_2)\rangle - \langle I(t)\rangle^2]. \quad (1)$$

Here i(t) is the photocurrent, I(t) is the light intensity, and T is the response time of the photodetector. According to the extended Huygens-Fresnel principle[1], the optical field at the receiver plane can be expressed in terms of the integral optical field at an intermediate plane:

$$E(\vec{r},L,t) \sim \iint_\Sigma d^2s E(\vec{s},z,t) \exp\left[\frac{ik|\vec{s}-\vec{r}|}{2(L-z)} + i\Psi(\vec{s},\vec{r},t)\right], \quad (2)$$

where $\Psi(\vec{s},\vec{r})$ is the complex phase of the wave propagating through the turbulent medium from the point (s,z) to the point (r,L). As follows from expression (2), the values of the averaging in expression (1) are of fourth order in the field moment:

$$\langle E(\vec{s}_1,z,t_1)E^*(\vec{s}_2,z,t_1)E(\vec{s}_3,z,t_2)E^*(\vec{s}_4,z,t_2)\rangle. \quad (3)$$

For Gaussian statistics, this fourth order moment can be expressed in terms of the second order moments:

$$\langle E(\vec{s}_1,z,t_1)E^*(\vec{s}_2,z,t_1)E(\vec{s}_3,z,t_2)E^*(\vec{s}_4,z,t_2)\rangle =$$
$$\langle E(\vec{s}_1,z,t_1)E^*(\vec{s}_2,z,t_1)\rangle\langle E(\vec{s}_3,z,t_2)E^*(\vec{s}_4,z,t_2)\rangle + \langle E(\vec{s}_1,z,t_1)E^*(\vec{s}_4,z,t_2)\rangle\langle E(\vec{s}_3,z,t_2)E^*(\vec{s}_2,z,t_1)\rangle. \quad (4)$$

The typical difference between the times, $t_1$ and $t_2$, in (1) can be estimated as $|t_1 - t_2| \sim T$. If the response time of the photodetector, T, exceeds the average correlation time between two coherent spots $\tau_c$, $T \gg \tau_c$, the second term on the right-hand side of the expression (4) is equal to zero. As a result, from Eq. (1) we obtain $\langle I(t_1)I(t_2)\rangle = \langle I\rangle^2$. This shows that the scintillation index $\sigma^2 = \left(\langle I^2 \rangle - \langle I \rangle^2\right)/\langle I \rangle^2$ is equal to zero. In the opposite case, when the correlation time is much longer than the photodetector response time, $T \ll \tau_c$, the second term in the expression (4) is equal to the first term, and the scintillation index is equal to unity.

As the above considerations show, in order to exploit the unique properties of Gaussian statistics, the time response of the photodetector must be much longer than the inverse frame rate of the SLM. Another requirement is that the number of individual coherent spots in the initial beam must be sufficiently large. In other words, the coherence radius, $r_c$, must be much smaller than the beam radius, $r_0$. Note that the minimum size of the initial coherence radius, $r_c$, is limited by two physical effects. First, the angular spreading of the laser beam is defined by the diffraction angle, $\theta \sim \lambda/r_c$. Consequently, for a very small coherent radius, $r_c$, the beam spread will be unacceptably large. Second, for a very small initial coherence radius, the diffraction effect will dominate in the formation of the beam coherence in comparison with the influence of the atmospheric turbulence. In this case, according to the Van Cittert-Zernike theorem[15], the coherence radius will increase during the propagation. Thus, a very important requirement can be formulated: In order to significantly suppress the laser beam scintillations, one must work in the regime in which an optimal initial coherence radius, $r_c$, is chosen which satisfies the inequality, $r_{min} < r_c < r_{max}$. The optimal initial coherence radius, $r_c$, depends on the strength of the atmospheric turbulence and the propagation length, L. The adaptive control of the initial coherence can be achieved using a feedback channel. Either a *rf* or an optical channel could provide feedback from the measurements of the scintillation index at the receiver to the SLM at the laser source.

**3. Calculation of the Scintillation Index for the Case of Strong Turbulence**

Our analysis is based on the equation for the fourth-order correlation function derived by Tatarskii in the Markov approximation.[16,17]. The equation for the correlation function

$$\Gamma_4(\zeta; \vec{\rho}_1, \vec{\rho}_1^{\,'}, \vec{\rho}_2, \vec{\rho}_2^{\,'}) = \langle E(\zeta, \vec{\rho}_1) E^*(\zeta, \vec{\rho}_1^{\,'}) E(\zeta, \vec{\rho}_2) E^*(\zeta, \vec{\rho}_2^{\,'}) \rangle \qquad (5)$$

has the form

$$\frac{\partial \Gamma_4}{\partial \zeta} = \frac{i}{2q}(\Delta_1 + \Delta_2 - \Delta_1' - \Delta_2')\Gamma_4 - F(\zeta; \vec{\rho}_1, \vec{\rho}_1', \vec{\rho}_2, \vec{\rho}_2')\Gamma_4, \tag{6}$$

where $\zeta = x/L, \vec{\rho}_{1,2} = \vec{r}_{1,2}/\rho_0, \vec{\rho}_{1,2}' = \vec{r}_{1,2}'/\rho_0$ ($x$ is the longitudinal coordinate), $\vec{r}_{1,2}, \vec{r}_{1,2}'$ are the transversal coordinates, $L$ is the propagation length, $\rho_0$ is the normalizing transverse scale, which is chosen below, $q = \frac{k\rho_0^2}{L}$, where $k$ is the wave number.

$$F(\zeta, \vec{r}_1, \vec{r}_2, \vec{\rho}) = H(\zeta, \vec{r}_1 + \vec{\rho}/2) + H(\zeta, \vec{r}_1 - \vec{\rho}/2) + H(\zeta, \vec{r}_2 + \vec{\rho}/2) + H(\zeta, \vec{r}_2 - \vec{\rho}/2) \\ - H(\zeta, \vec{r}_1 + \vec{r}_2) - H(\zeta, \vec{r}_1 - \vec{r}_2). \tag{7}$$

In the expression (7) we introduced the new variables

$$\vec{r}_1 = \frac{1}{2}(\vec{\rho}_1 - \vec{\rho}_2 + \vec{\rho}_1' - \vec{\rho}_2'); \vec{r}_2 = \frac{1}{2}(\vec{\rho}_1 - \vec{\rho}_2 - \vec{\rho}_1' + \vec{\rho}_2'), \\ \vec{\rho} = \vec{\rho}_1 + \vec{\rho}_2 - \vec{\rho}_1' - \vec{\rho}_2'; R = \frac{1}{4}(\vec{\rho}_1 + \vec{\rho}_2 + \vec{\rho}_1' + \vec{\rho}_2'). \tag{8}$$

In these new variables, Eq. (6) takes the form

$$\frac{\partial \Gamma_4}{\partial \zeta} = \frac{i}{q}(\nabla_R \nabla_\rho + \nabla_{r_1}\nabla_{r_2})\Gamma_4 - F(\zeta, \vec{r}_1, \vec{r}_2, \vec{\rho})\Gamma_4, \tag{9}$$

where

$$H(\zeta, \vec{\rho}) = 8\iint \Phi_n(\zeta, \vec{\kappa})[1 - \cos\vec{\kappa}(\vec{\rho}_1 - \vec{\rho}_2)]d^2\kappa, \tag{10}$$

and $\Phi_n(\zeta, \vec{\kappa})$ is the spectral density of the structure function of the refractive index, which is given by

$$\langle \delta n(\zeta, \vec{\rho})\delta n(\zeta, \vec{\rho}')\rangle = 2\pi\delta(\zeta - \zeta')\iint \Phi_n(\zeta, \vec{\kappa})\exp(-\vec{\kappa}(\vec{\rho} - \vec{\rho}'))d^2\kappa. \tag{11}$$

Following Tatarskii[16], we chose the spectral density of the structure function of the refractive index in the form

$$\Phi_n(\zeta,\vec{\kappa}) = 0.033 C_n^2 \kappa^{-11/3} \exp\left(-\frac{\kappa^2}{\kappa_m^2}\right). \tag{12}$$

In this case the analytical approximations for the function $H(\zeta,\vec{\rho})$ have the form

$$H(\zeta,\vec{\rho}) = \begin{cases} 1.64 C_n^2 k^2 \rho_0^2 \rho^2 l_0^{-1/3}, & \text{for } \rho \ll \dfrac{l_0}{\rho_0}, \\ 1.24 C_n^2 k^2 \rho_0^{5/3} \rho^{5/3}, & \text{for } \rho \gg \dfrac{l_0}{\rho_0}, \end{cases} \tag{13}$$

where $l_0 = 5.92 \kappa_m$. The transverse scale, $\rho_0$, is the scale of variation of the phase structure function of the plane waves corresponding to the path L. It is defined by the equation[15]: $1.64 C_n^2 k^2 L \rho_0^2 l_0^{-1/3} = 1$.

According to the procedure discussed in Section 2, we assume that the light source emits a partially coherent light with the Gaussian statistics at the source plane, $\zeta = 0$. Hence, the fourth-order correlation function can be expressed in terms of the second order correlation functions

$$\Gamma_{4,0}(\vec{\rho}_1, \vec{\rho}_1', \vec{\rho}_2, \vec{\rho}_2') = \langle E(\zeta=0, \vec{\rho}_1) E^*(\zeta=0, \vec{\rho}_1') \rangle \langle E(\zeta=0, \vec{\rho}_2) E^*(\zeta=0, \vec{\rho}_2') \rangle + \\ \langle E(\zeta=0, \vec{\rho}_1) E^*(\zeta=0, \vec{\rho}_2') \rangle \langle E(\zeta=0, \vec{\rho}_2) E^*(\zeta=0, \vec{\rho}_1') \rangle, \tag{14}$$

where the normalized second order correlation function is given by the expression

$$\Gamma_{2,0}(\vec{\rho}_{1,2}, \vec{\rho}_{1,2}') = \langle E(\zeta=0, \vec{\rho}_{1,2}) E^*(\zeta=0, \vec{\rho}_{1,2}') \rangle = \exp\left(-\frac{\vec{\rho}_{1,2}^2 + \vec{\rho}_{1,2}'^2}{2 r_0^2}\right) \exp\left(-\frac{(\vec{\rho}_{1,2} - \vec{\rho}_{1,2}')^2}{r_c^2}\right). \tag{15}$$

In Eq. (15), $r_0$ is the beam radius and $r_c$ is the coherence radius. We consider the case of small coherence radii in comparison with the beam radius: $r_c \ll r_0$. The conventional approach to the problem of laser beam propagation is based on the assumption of small deviations of the beam parameters from those which correspond to free-space propagation.[1] This approach is limited to the conditions of weak turbulence or short propagation lengths. Another approach was developed by Yakushkin.[10] His approach is based on the fact that for any relatively long distance, the coherence radius is smaller than the beam radius. Starting with the exact solution of Eq. (9) for a beam with $r_c = 0$, Yakushkin developed a perturbation theory in which the small parameter, $r_c/r_0 \ll 1,$ is the ratio of the coherence radius $r_c$ to the beam radius $r_0$. In[10] the case of an initially fully coherent beam was considered. Thus, his theory was actually an asymptotic theory, applicable to relatively long distances. In our case, we have initially a partially coherent beam.

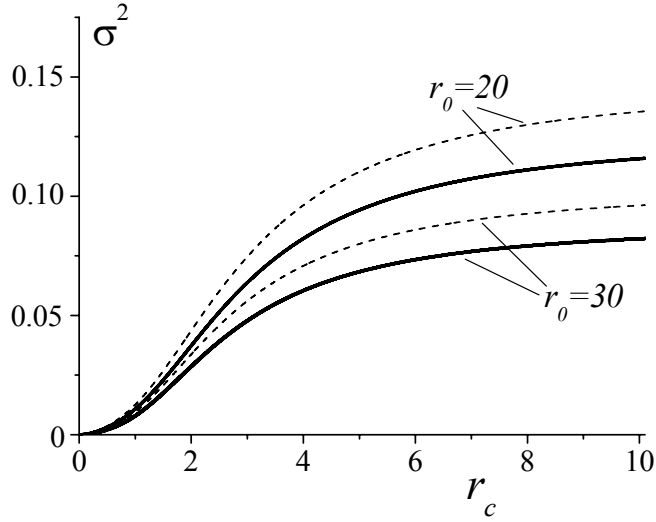

Fig. 1. The dependence of the scintillation index on the initial coherence radius (for the center of the beam). In the dimensionless quantities $r_{c,0}$ the unit is $\rho_0 = [1.64 C_n^2 k^2 L l_0^{-1/3}]^{-1/2}$. The following values of the parameters were used: $L = 3.7 km$, $C_n^2 = 10^{-13} m^{-2/3}$ ( the dashes lines), $L = 10 km$, $C_n^2 = 1.4 \times 10^{-14} m^{-2/3}$ (the solid lines), $\lambda = 1.55 \mu m$, $l_0 = 2 \times 10^{-2} m$.

Hence our approach is applicable to any distance. In[11] the approach[10] was applied to the case of an initially partially coherent beam. However, the authors of[11] did not discuss the details of the intermediate approximations. Although, their results are qualitatively similar to ours, the final expressions are different. In particular, we obtained a much simpler expression for the scintillation index. (See below the expression (21).) Authors[12] solved the quantum kinetic equation for the photon distribution function. Their results are also qualitatively similar to ours. In [13] the author considered an incoherent source, which corresponds to the case of the coherence radius equal to zero in our consideration. The scintillation index in this case is equal to zero (see Fig. 1), what corresponds to the case of infinitely strong turbulence in [13]. The formula (21) describes the case of finite coherence radius, when the scintillation index is different from zero even in the case of very strong turbulence.

The solution of Eq. (9) can be written in the integral form

$$\Gamma_4(\vec{P},\zeta) = \int \Gamma_{4,0}(\vec{P}')G(\vec{P},\vec{P}',\zeta)d\vec{P}', \tag{16}$$

where $\vec{P}$ is the set $\vec{r}_1, \vec{r}_2, \vec{\rho}, \vec{R}$, and $G(\vec{P},\vec{P}',\zeta)$ is the Green's function. We assume that the main contribution to the integral is due to areas of coherence, in which the difference between two vectors has the value $|\vec{\rho}_{1,2} - \vec{\rho}'_{1,2}| \sim \rho_c$, where $\rho_c$ is the coherence radius for the propagation length $\zeta$. Note that initially $(\zeta = 0)$, the coherence radius is defined by the source $\rho_c = r_c$. For these areas the values of the vectors $\vec{r}_2$ and $\vec{\rho}$ are of order $\rho_c$. In this case, the values of the vector $\vec{r}_1$ are of the order of the beam radius $\rho_0$, which at the source plane is equal to the initial beam radius $r_0$. We assume that the coherence radius is much smaller than the beam radius. Hence we have $r_2, \rho \ll r_1$. Since the function $\Gamma_4$, must be symmetric with respect to $r_1$ and $r_2$, there are other areas where $r_2 \gg r_1, \rho$. Taking into account these inequalities we can obtain the zeroth order approximation for the function $F(\zeta,\vec{r}_1,\vec{r}_2,\vec{\rho})$:

$$F^0(\zeta,\vec{r}_1,\vec{r}_2,\vec{\rho}) = H(\zeta,\vec{r}_2 + \vec{\rho}/2) + H(\zeta,\vec{r}_2 - \vec{\rho}/2). \tag{17}$$

In this approximation the Green's function has the form

$$G^0(\vec{P},\vec{P}',\zeta) = \frac{q^4}{16\pi^4\zeta^2}\exp\left\{\frac{iq}{\zeta}[(\vec{R}-\vec{R}')(\vec{\rho}-\vec{\rho}') + (\vec{r}_1-\vec{r}_1')(\vec{r}_2-\vec{r}_2')]\right\} \\ \times \exp\left\{-\zeta(\vec{r}_2-\vec{r}_2' + (\vec{\rho}-\vec{\rho}')/2)^2 - \zeta(\vec{r}_2-\vec{r}_2' - (\vec{\rho}-\vec{\rho}')/2)^2\right\}. \tag{18}$$

In (18), we use the quadratic approximation for the function $H(\zeta,\vec{\rho})$ (the upper line of expression (13)). Writing the Green's function in the form $G = G^0 + \Delta G$, and using the Green's function formula, we obtain the following integral equation:

$$G(\vec{P},\vec{P}',\zeta) = G^0(\vec{P},\vec{P}',\zeta) - \int_0^\zeta d\zeta' \int \Delta F G^0(\vec{P}'',\vec{P}',\zeta') G(\vec{P},\vec{P}'',\zeta-\zeta') d\vec{P}'' d\zeta', \qquad (19)$$

where $\Delta F = F - F^0$. Using (16) and (19), we obtain the integral equation for $\Gamma_4$, (which is equivalent to the differential equation (9)):

$$\Gamma_4(\vec{P},\zeta) = \int \Gamma_{4,0}(\vec{P}') G^0(\vec{P}',\vec{P},\zeta) d\vec{P}' - \int_0^\zeta d\zeta' \Delta F G^0(\vec{P},\vec{P}',\zeta') \Gamma_4(\vec{P}',\zeta-\zeta') dP'. \qquad (20)$$

We use Eq. (20) in our perturbation theory. The zeroth approximation for $\Gamma_4$ is given by the first term on the right hand side of Eq. (20). The first-order approximation term is obtained by substituting the zeroth approximation term in the second term on the right hand side of equation (20). To calculate the term $\Delta F$, we have to use Taylor serious expansion for the coefficients $H(\zeta,\vec{r}_1 \pm \vec{\rho}/2), H(\zeta,\vec{r}_1 \pm \vec{r}_2)$ in Eq. (7). In the first and second terms of the expansion we have to use approximation presented by the lower line in Eq. (13) because the beam radius $\rho_0 \sim r_1$ is assumed to be much larger than the smallest turbulence scale $l_0$, $\rho_0 \gg l_0$. Performing straightforward but cumbersome calculations, we obtain the following expression for the scintillation index (at the center of the beam)

$$\sigma^2 = 0.68\alpha_0(\xi=1)Q^{1/6} q^{1/6} \int_0^1 d\xi \frac{\left[\dfrac{0.286}{\beta(\xi)} - \dfrac{0.43}{\gamma(\xi)} + \dfrac{0.157\mu^2(\xi)}{\gamma^2(\xi)\alpha_2(\xi)}\right]}{(1-\xi^2)\alpha_0(\xi)\gamma(\xi)\beta(\xi)\alpha_2^{5/6}(\xi)}, \qquad (21)$$

where

$$Q = \frac{kl_0^2}{L}; \quad \alpha_0 = \xi + \frac{1}{r_f^2} + \frac{q^2 r_0^2}{4\xi^2}; \quad \frac{1}{r_f^2} = \frac{1}{r_c^2} + \frac{1}{4r_0^2}; \quad \mu = \frac{1}{1-\xi} + \frac{1}{2\xi\alpha_0 r_f^2}; \quad \nu = \frac{2}{\alpha_0 r_f^2}\left(\xi + \frac{q^2 r_0^2}{4\xi^2}\right);$$

$$\beta = \nu + 1 - \xi + \frac{\alpha_0 \xi^2 \mu^2}{2}; \quad \gamma = \nu + 2(1-\xi); \quad \alpha_2 = \frac{\mu^2}{4\gamma} + \frac{1}{8\xi^2 \alpha_0}.$$

The scintillation index decreases as the initial coherence radius, $r_c$, decreases, as can be seen in Fig. 1. For the coherence radii less than 4, the scintillation index has a quadratic-like dependence. For a certain value of the coherence radius, a larger beam radius corresponds to a smaller scintillation index. Actually, the scintillation index decreases linearly with an inverse number of coherent spots $\sigma^2 \sim N_c^{-1} = r_c^2/r_0^2$. Thus the scintillation index decreases by an order of magnitude as the coherence radius $r_c$ decreases from 3 to 1 (see Fig. 1).

## 4. Design of an Optical System Based on Spectral Amplitude Encoding of a Broad Band Pulsed Laser

We propose to encode digital data in the spectrum of a wide-band source such as Ti: sapphire laser. We assume that this laser operates at a high repetition rate. Usually Ti: sapphire lasers can operate at a repetition rate in a broad range from a few Hz up to GHz. If each series of N laser pulses (the number of pulses depends on the averaging time of the photosensor) contains kilobits of data and the series repetition rate is several MHz, then the data rate is a few Gbps. Usually, information is encoded as an amplitude modulation in time of a continuous wave laser beam. In this case, intensity fluctuations make significant contributions to the BER. Our spectral domain encoding technique is less sensitive to intensity and phase fluctuations because the information is decoded in a massively parallel way using a relatively slow photosensor, which minimizes the scintillations by time averaging. Spectral-domain encoding can be achieved using a wide-band Ti: sapphire laser with an electro-optical SLM, as demonstrated in Fig. 2. The spectrum of each laser pulse is dispersed spatially and then encoded by the SLM whose pixels can be turned on and off. The light traversing the SLM will have certain spectral bins turned on or off, depending on whether the corresponding pixels of the SLM are on or off. In our approach, the spatial coherence of the initial beam is also formed by the SLM. The encoded signal is sent through a second SLM which modifies the transverse coherence of the beam. The optimal value of the radius of coherence, $r_c$, is maintained by using a feedback loop between the SLM and the photosensor (not shown). At the receiver, the wavelength-modulated signal is dispersed in an optical spectrometer, *i.e.* a monochromator, and detected by a high-speed charge-coupled device (CCD). The electronic signal from the CCD is then processed by a high-speed data processing unit. Note, that in our method the spectral grating spreads the spectrum of the laser pulse along a single coordinate. Therefore we need a single array LSM for encoding and a single array CCD for sensing.

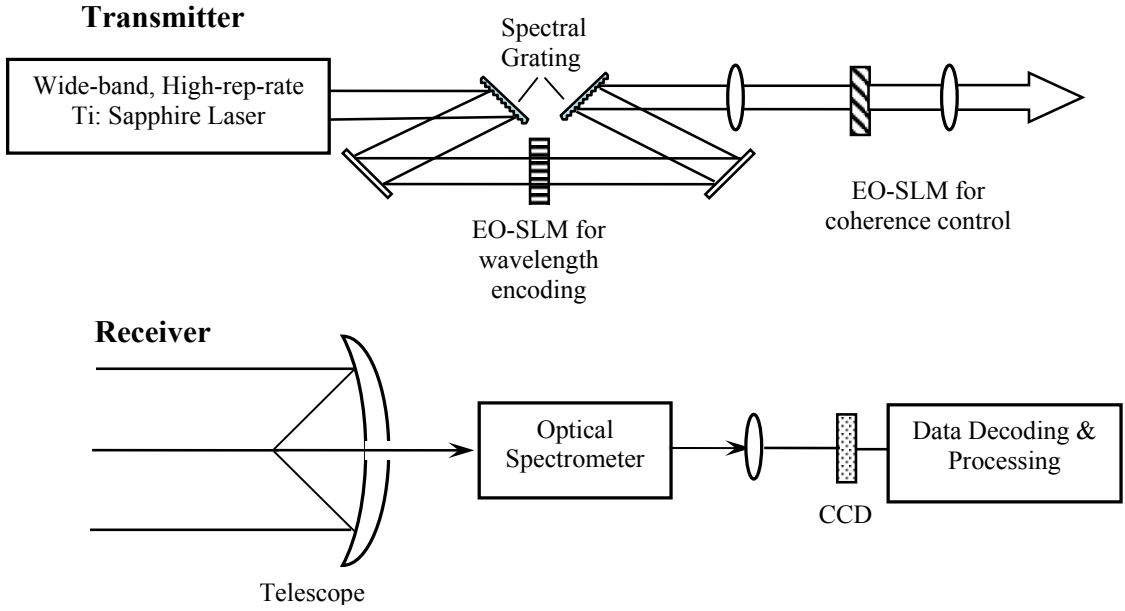

Fig. 2. Schematic for wide-band laser communication with wavelength encoding.

Let us estimate the values of parameters needed to achieve Gbps rate. The grating dispersion is described by the grating equation

$$\sin\theta_{inc} - \sin\theta_{diff} = m\frac{\lambda}{d}, \qquad (22)$$

where $\theta_{inc}$ is the incident angle, $\theta_{diff}$ is the diffraction angle, $m$ is the diffraction order, and $d$ is the grating period. Usually, the incident angle is fixed $\theta_{inc} = const$. Then the angle variation $\delta\theta_{diff}$ as a function of the wavelength variation $\delta\lambda$ is

$$\delta\theta_{diff} = m\frac{\delta\lambda}{d\cos\theta_{diff}}. \qquad (23)$$

The spatial image at the SLM plane of the spectral interval, corresponding to the distance between two neighboring bits, must exceed the size of individual pixel of the SLM, $l_{pxl}$. Then $\delta\theta_{diff} F \geq l_{pxl}$, where $F$ is the focal distance of the imaging lens. From the expression (23) we obtain the following estimate of the spectral interval

$$\delta\lambda = \frac{l_{pxl} d}{Fm}. \quad (24)$$

Using (24), we estimate the information capacity, $M$, of the laser pulse which has the spectral width $\Delta\lambda$,

$$M = \frac{\Delta\lambda}{\delta\lambda} = \frac{\Delta\lambda F m}{l_{pxl} d \cos\theta_{diff}} \quad (25)$$

For the values of parameters $\Delta\lambda = 40 nm$ (for the wavelength $\lambda = 1.55 \mu m$ this spectral width corresponds to a pulse duration 200 fs), $F = 10 cm$, $m = 2$, $l_{pxl} = 10 \mu m$, $d^{-1} = 1.5 \times 10^3 mm^{-1}$, $\cos\theta_{diff} = 0.6$, we obtain $N = 2 \times 10^3$. Using a commercially available CCD array with a frame rate of 1 MHz, a single array MQW encoding SLM with the frame rate of 1 MHz, 10×10 pixels MQW SLM for the coherence control with the frame rate of 10 MHz, and Ti: sapphire laser with the pulse duration of 200 fs and the repetition rate 10 MHz, we achieve the data rate of 2 Gigabits per second.

**5. Utilization of LEDs for Short-Distance FSOC**

The problem of low-cost, high-speed connections of individual customers with backbone optical fiber channels is known as the "last mile bottleneck" problem. The increasing demand for wireless connection to optical fiber channel stimulated needs of FSOC technology for distances of 0.25–1mile. Being a rapid deployment system, short-distance FSCO can be used also for disaster recovery following natural catastrophes and terrorist attacks[18]. Until recently, FSOC used lasers as light sources. The LED as a source was considered as undesirable for three reasons: i) limited modulation frequency (typically up to 100 Mbps), ii) high radiation divergence, and iii) low power. However, when compared with lasers for short-distance FSOC, LEDs have the advantages of higher reliability, reduced temperature sensitivity, immunity to optical feedback, and lower cost. For these reasons, LEDs are utilized for short distance FSOC with moderate data rates limited by 100 Mbps.[18]

Since LEDs are a source of spontaneous non-coherent emission, their emission has Gaussian statistics with a very short coherence time, of the order $10^{-13}$ s or less. Contrary to the conventional time domain encoding technology, the spectral encoding technology does not require high modulation rates: A modulation rate of about several MHz can provide a channel with a Gbps data rate.

## 6. Conclusion

We have presented the new concept of a free space, a few Gbps speed optical communication system based on spectral encoding of radiation from a broadband pulsed laser. We have shown that, in combination with control of the partial coherence of the laser beam and the relatively slow photosensor, scintillations can be suppressed by orders of magnitude for communication distances beyond 10 km. Also, we considered the spectral encoding of radiation of a LED as a gigabit rate solution of the "last mile" problem and a rapid deployment system for disaster recovery.

## Acknowledgement

The authors would like to thank A.A. Chumack for many stimulating discussions. This work was carried out under the auspices of the National Nuclear Security Administration of the U.S. Department of Energy at Los Alamos National Laboratory under Contract No. DE-AC52-06NA25396.